\documentclass[preprint,amsmath,amssymb,aps,showkeys,showpacs]{revtex4}
\usepackage[colorlinks=true, pdfstartview=FitV, linkcolor=blue, citecolor=red, urlcolor=magenta]{hyperref}
\usepackage{graphicx}
\usepackage{latexsym}
\usepackage{amsmath}
\usepackage{amsfonts}
\usepackage{amssymb}
\usepackage{verbatim}
\usepackage{dcolumn}
\usepackage{amssymb}
\usepackage{bm}
\usepackage{float}
\usepackage{subfigure}
\usepackage[english]{babel}
\newcommand{\be}{\begin{equation}}
\newcommand{\ee}{\end{equation}}
\newcommand{\bea}{\begin{eqnarray}}
\newcommand{\eea}{\end{eqnarray}}

\begin{document}

\newcommand{\NITK}{
\affiliation{Department of Physics, National Institute of Technology Karnataka, Surathkal  575 025, India}
}

\newcommand{\IIT}{\affiliation{
Department of Physics, Indian Institute of Technology, Ropar, Rupnagar, Punjab 140 001, India
}}

\title{Microstructure and continuous phase transition
of a regular Hayward black hole in anti-de Sitter
spacetime}

\author{Naveena Kumara A.}
\email{naviphysics@gmail.com}
\NITK
\author{Ahmed Rizwan C.L.}
\email{ahmedrizwancl@gmail.com}
\NITK
\author{Kartheek Hegde}
\email{hegde.kartheek@gmail.com}
\NITK
\author{Ajith K.M.}
\email{ajithkm@gmail.com}
\NITK
\author{Md Sabir Ali}
\email{alimd.sabir3@gmail.com}
\IIT

%

\begin{abstract}
We study the phase transition of a regular Hayward-AdS black hole by introducing a new order parameter, the potential conjugate to the magnetic charge due to the non-linearly coupled electromagnetic field. We use Landau continuous phase transition theory to discuss the van der Waals like critical phenomena of the black hole. The popular interpretation of the AdS black hole phase transition as between a large and a small black hole is reinterpreted as the transition between a high potential phase and a low potential phase. The underlying microstructure for this phase transition is probed using the Ruppeiner geometry. By investigating the behaviour of the Ruppeiner scalar curvature, we find that the charged and uncharged (effective) molecules of the black hole have distinct microstructures analogous to fermion and boson gas. 

\end{abstract}

\keywords{}

\maketitle


\section{Introduction}
In black hole physics one of the most intriguing question, since it's prediction, is that does it have a microscopic structure? The simplest answer to this as said by Boltzmann ``If you can heat it, it has microscopic structure". In the pioneering work of Hawking and Bekenstein \citep{Hawking:1974sw, Bekenstein1972, Bekenstein1973, Bardeen1973} it was shown that a black hole is not only a gravity system but also a thermal system. A black hole possesses temperature and entropy which are related to its surface gravity and event horizon area, respectively. The Hawking temperature of the black hole will change during the absorption and emission of matter, which suggests that the black hole must have a microstructure. Inspired by this, several approaches were developed to probe the microscopic origin of black hole thermodynamics \citep{Wei2015, Wei2019a, Wei2019b, Guo2019, Miao2017, Zangeneh2017, Wei:2019ctz, Xu:2019nnp, Chabab2018, Deng2017, Miao2019a, Chen2019, Du2019, Dehyadegari2017}. Unlike an ordinary thermodynamic system, where the macroscopic quantities are constructed from the microscopic knowledge of the system, the statistical investigation of the black hole is carried out in a reverse order. The microstructure of the black hole is scrutinized from its macroscopic thermodynamic quantities by studying the phase transitions.

Black holes exhibit wide variety of phase transitions like those of ordinary thermodynamic systems. Particularly, phase transitions in AdS background have been extensively studied in recent years. The most important assumption in these studies is the identification of the cosmological constant as the thermodynamic variable pressure \citep{Kastor:2009wy}. For a charged AdS black hole the first order (discontinuous) and second-order (continuous) phase transition features are analytically similar to van der Waals (vdW) fluid \citep{Kubiznak2012, Gunasekaran2012, Kubiznak:2016qmn}. By studying these phase transitions, the microstructure of the black hole can be investigated via thermodynamic geometry methods. Ruppeiner geometry has proven to be quite useful and interesting to probe the interactions of a black hole from its macroscopic thermodynamic properties \citep{Ruppeinerb2008}. In this technique, a line element is defined in a parametric space, which is the measure of the distance between two neighbouring fluctuation states \citep{Ruppeiner95}. The curvature scalar constructed out of this line element is an indicator for the nature of the constituents of the system, positive sign for repulsive and negative sign for attractive interactions. The interpretation of this correspondence in black hole physics is adopted from the results obtained by the the applications of Ruppeiner geometry in ordinary thermodynamic systems \citep{Janyszek_1990, Oshima_1999}.

Choosing a parameter space with coordinates as the mass and the pressure for a charged AdS black hole, the Ruppeiner curvature scalar can be written in terms of the molecular density of the black hole microstates \citep{Wei2015}. By introducing the concept of black hole molecule, the authors have studied the phase transition and the interaction between the black hole molecules in two distinct phases. The molecular number density measures the microscopic degrees of the freedom, using which the order parameter is constructed. However, recently it is shown that for RN-AdS black hole the phase transition is regulated by the electric potential \citep{Guo2019}. The black hole can exist in any of the three potential phases, namely, high potential phase, low potential phase and neutral potential phase. The neutral potential of the black hole is analogous to the liquid-vapour coexistence phase in a vdW system. The potential due to the charge $Q$ serves as the order parameter. The microscopic and phase transition study is carried out by using the Landau continuous phase transition theory. The influence of charge, the key entity in phase transition as it was conjectured,  on the microstructure is investigated via Ruppeiner geometry. In this regard, one of the motivation for our research stems from the curiosity about the phase structure of the black holes which are composed of magnetic charges.

It is widely believed that due to the imperfections in classical theories of gravity there exists spacetime singularities which may be remedied by considering quantum effects, which calls for a quantum theory of gravity. Since there is no conclusive theory of quantum gravity, the intuitive answers to these are given on semi-classical regime. The first regular black hole model, which is free from the singularity, was proposed by Bardeen \citep{bardeen1968non}, where the singularity is replaced with a de Sitter core. Subsequent studies showed that a regular black hole can be realized as a solution of Einstein gravity coupled to non-linear electrodynamics source, which is the charged version of Bardeen black hole \citep{AyonBeato:1998ub, AyonBeato:2000zs}. A regular black hole can even be interpreted as the gravitational field of a non-linear magnetic monopole. A different kind of regular solution was provided by Hayward \citep{Hayward:2005gi}. Similar to Bardeen solution it is a degenerate configuration
of the gravitational field of a non-linear magnetic monopole. The solution carries magnetic charges and
a free integration constant. 

It is more reasonable to rephrase the question we posed in the beginning as, does a black hole without singularity have a microstructure? In this work, we investigate the phase structure of regular Hayward black hole in AdS background. The article is organized as follows. In section \ref{secone} the phase transition is studied using Landau theory of continuous phase transition theory followed by microstructure probe using Ruppeiner geometry in section \ref{sectwo}. Results are presented in section \ref{secthree}.


\section{Phase Transition of Regular Hayward Black Hole}
\label{secone}

\subsection{Thermodynamics of the Black Hole}
The action of the Einstein gravity coupled to a nonlinear electromagnetic field in the background of AdS spacetime is \citep{Fan:2016hvf, Fan:2016rih}

\begin{equation}
I=\frac{1}{16\pi}\int d^4x\sqrt{-g}(R-2\Lambda -\mathcal{L}).
\end{equation}
In the above expression $F=dA$ is the field strength of the vector field $\mathcal{F}=F^{\mu \nu}F_{\mu \nu}$ and the Lagrangian density $\mathcal{L}$ is a function of field strength. The Hayward class of solutions arise from the Lagrangian density,
\begin{equation}
\mathcal{L}=\frac{4 \mu}{\alpha}\frac{(\alpha \mathcal{F})^{(\mu +3 )/4}}{(1+(\alpha \mathcal{F})^{\mu /4})^2},
\end{equation}
where $\mu >0$ is a dimensionless constant and $\alpha>0$ has the dimension of length squared. The general static spherically symmetric solution is given by,
\begin{equation}
ds^2=-f(r)dt^2+\frac{dr^2}{f(r)}+r^2d\Omega ^2, \quad \text{with} \quad A=Q_m \cos \theta d \phi ,
\end{equation} 
where $d\Omega ^2=d\theta ^2+\sin \theta ^2d\phi ^2$ and $Q_m$ is the total magnetic charge of the black hole
\begin{equation}
Q_m=\frac{1}{4\pi} \int _{\Sigma _2}F.
\end{equation}
In the AdS background the function $f(r)$ has the following form,
\begin{equation}
f(r)=\frac{r^2}{l^2}+1-\frac{2M_S}{r}-\frac{2\alpha ^{-1}g^3r^{\mu -1}}{r^\mu +g^\mu}
\end{equation}
where $M_S$ is the Schwarzchild mass and $g$ is the free integration constant which is related to magnetic charge as,
\begin{equation}
Q_m=\frac{g^2}{\sqrt{2\alpha}}.
\end{equation}
The ADM mass can be read off from the asymptotic behaviour of the function $f(r)$ and it has the following form,
\begin{equation}
M_{ADM}=M_S+M_{em} , \quad M_{em}=\alpha ^{-1}g^3.
\end{equation}
The Schwarzchild mass $(M_S)$ is the resultant of the nonlinear self interactions of the massless graviton. The other contribution $M_{em}$ is associated with the nonlinear interaction between the graviton and the photon. The Hayward black hole solution corresponds to the case $M_S=0$ and $\mu=3$ which has the following metric function (henceforth we will take $M_{em}=M$),
\begin{equation}
f(r)=1-\frac{2 M r^2}{g^3+r^3}+\frac{8}{3} \pi  P r^2.
\end{equation}
The pressure $P$ is related to the cosmological constant $\Lambda$ as $P=-
\Lambda /8\pi$.
The event horizon of this black hole is determined by the condition $f(r_+)=0$, using which the explicit expression for mass $M$ can be written. The Hawking temperature  of the black hole which is related to the surface gravity $\kappa$, is obtained as
\begin{equation}
T=\frac{f'(r_+)}{4\pi}=\frac{2 P r^4}{g^3+r^3}-\frac{g^3}{2 \pi  r \left(g^3+r^3\right)}+\frac{r^2}{4 \pi  \left(g^3+r^3\right)}. \label{temperature}
\end{equation}
With this temperature we can write the first law of thermodynamics in the conventional form,
\begin{equation}
dM=TdS+\Psi dQ_m+VdP+\Pi d \alpha.
\end{equation}
Where $\Psi$ is the conjugate potential for the magnetic charge $Q_m$, $\Pi$ is conjugate to the parameter $\alpha$. The volume and entropy of the black hole have the following non trivial profile,
\begin{equation}
V=\frac{4}{3} \pi  \left(g^3+r^3\right) \quad \text{and} \quad S=2 \pi  \left(\frac{r^2}{2}-\frac{g^3}{r}\right).
\end{equation}
The Potential reads as,
\begin{equation}
\Psi=\frac{3g^4(2r_+^3+g^3)}{\sqrt{2\alpha}(r_+^3+g^3)^2}.
\end{equation}
Rearranging the expression \ref{temperature} we have the equation of state,
\begin{equation}
P=\frac{g^3}{4 \pi  r^5}+\frac{g^3 T}{2 r^4}-\frac{1}{8 \pi  r^2}+\frac{T}{2 r}.
\end{equation}
The critical behaviour of regular Hayward black hole can be studied choosing a pair of conjugate variables like $(P-V)$ or $(T-S)$. Choosing the pair $(P,V)$ we have the Maxwell's equal area law in the following form,
\begin{equation}
P_0(V_2-V_1)=\int _{V_1}^{V_2}PdV. \label{equalarea}
\end{equation}
With equation \ref{equalarea} and using the corresponding expressions for $P_0(V_1)$ and $P_0(V_2)$ from equation of state we obtain,
\begin{equation}
r_2=g\left[ \frac{x \left(x^3+6 x^2+6 x+1\right)+\sqrt{y}}{x^4}\right]^{1/3}, \label{r2eqn}
\end{equation}
\begin{equation}
P_0=\frac{3 \left[\frac{\sqrt{y}+ x \left(x^3+6 x^2+6 x+1\right)}{x^4}\right]^{1/3} \left[\left(-2 x^4-11 x^3-20 x^2-11 x-2\right) \sqrt{y}+ z\right]}{16 \pi  g^2 x \left(x^2+4 x+1\right) \left(3 x^2+4 x+3\right)^2}, \label{peqn}
\end{equation}
\begin{equation}
T_0=\frac{\left[\frac{\sqrt{y}+x \left(x^3+6 x^2+6 x+1\right)}{x^4}\right]^{2/3} \left[u-\left(x^3+4 x^2+4 x+1\right) \sqrt{y}\right]}{4 \pi  g x \left(3 x^4+16 x^3+22 x^2+16 x+3\right)}. \label{teqn}
\end{equation}

\begin{figure}[H]
\centering
\includegraphics[scale=1]{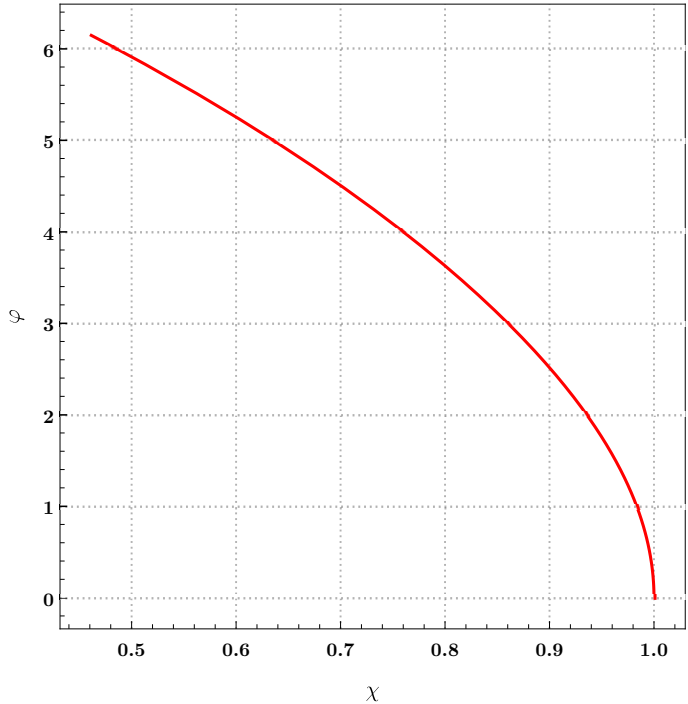}
\caption{The order parameter vs reduced temperature $(\varphi - \chi)$ plot of regular Hayward AdS black hole.}\label{ordercurve}
\end{figure}

Where
\begin{equation}
y=x^2 \left(x^6+12 x^5+54 x^4+82 x^3+54 x^2+12 x+1\right),
\end{equation}
\begin{equation}
z=x \left(2 x^7+23 x^6+104 x^5+213 x^4+213 x^3+104 x^2+23 x+2\right),
\end{equation}
\begin{equation}
u=x \left(x^6+10 x^5+37 x^4+54 x^3+37 x^2+10 x+1\right).
\end{equation}
We have taken $    x=r_1/r_2$, where $r_1$ and $r_2$ are the radii of black holes for first order phase transition points. $x=1$ gives the critical values of temperature $T$ and pressure $P$,
\begin{equation}
T_{c}=\frac{\left(5 \sqrt{2}-4 \sqrt{3}\right) \left(3 \sqrt{6}+7\right)^{2/3}}{4\ 2^{5/6} \pi  g}
\end{equation}
\begin{equation}
P_{c}=\frac{3 \left(\sqrt{6}+3\right)}{16\ 2^{2/3} \left(3 \sqrt{6}+7\right)^{5/3} \pi  g^2}.
\end{equation}
During the phase transition, a system undergoes a sudden change in its physical properties which is controlled by external thermodynamic variables.
A familiar example in day to day life is the solid-liquid-gas transition due to a change in the temperature or pressure. A common feature of these transitions is that the order or symmetry of the system changes at the transition point. At the phase transition point of the black hole there is a sudden change in the potential $\Psi$ which is controlled by pressure and temperature. This indicates the fact that the black hole microstructure are in different phases in different potentials. The two phases of the black hole with different symmetry and order are determined by the potentials,
\begin{equation}
\Psi _1=\frac{3g^4(2r_1^3+g^3)}{\sqrt{2\alpha}(r_1^3+g^3)^2} \quad \text{and} \quad
\Psi _2=\frac{3g^4(2r_2^3+g^3)}{\sqrt{2\alpha}(r_2^3+g^3)^2},
\end{equation}
where $r_1$ and $r_2$ can be written in terms of $x$ using expression \ref{r2eqn} and the critical value of the potential $\Psi _C$ can be obtained by setting $x=1$. We define the following order parameter to characterize the phase transition,
\begin{equation}
\varphi (T)= \frac{\Psi _1-\Psi _2}{\Psi _C}.
\end{equation}
The behaviour of the order parameter is shown in fig \ref{ordercurve}, where we have defined $\chi=T/T_C$. We examine the mechanism of black hole phase transition from the perspective of black hole magnetic charge, since the conjugate potential serves as order parameter.

\subsection{Phase Transition of the Black Hole}
In Landau theory, a  continuous phase transition is associated with a broken symmetry. In other words, the phase transition is the spontaneous breaking of this symmetry where the system chooses one state during this. Landau realized that an approximate form for the free energy can be constructed without prior knowledge about the microscopic states. In this theory, it is always possible to ascertain an order parameter which vanishes on the high-temperature side of the phase transition and has a nonzero value on the low-temperature side of the phase transition. That is, in one phase, the order parameter has a non-zero value, in another phase it vanishes. The order parameter describes the nature and extent of symmetry breaking. At the second-order phase transition, the order parameter grows continuously from the null value at the critical point. Since the order parameter approaches to zero at the phase transition, one can Taylor expand the free energy as a power series of the order parameter. The form of this expansion is governed by the symmetries of the theory. 

The phase transition in a black hole system can also be characterized by the symmetry and order degrees as in an ordinary thermodynamic system. In a black hole system, we can identify two phases with different potentials below the critical point (for $T<T_C$). The system in high potential phase $\varphi _1$ possess relatively ordered state with low symmetry. This corresponds to the certain orientation of black hole molecules under the action of the strong potential $\varphi _1$.  The system is in a low potential state $\varphi _2$ possess relatively low order degree and the high symmetry than the phase $\varphi _1$. This is due to the weakened orientation of the black hole molecules. The black hole molecules will have a certain orientation for all the values of temperature below critical value $T_C$. When the temperature is above a critical value ($T>T_C$) the thermal motion of the molecules increases and causes the black hole molecules to approach random orientation. The system now has higher symmetry than all the states below the critical temperature. For the less symmetrical phase (below the critical temperature with higher-order) the order parameter $\varphi$ is nonzero. For the more symmetrical phase (above critical temperature with lower order) the order parameter $\varphi$ is zero. 

Near the critical point $T_C$, the order parameter $\varphi$ is small and hence the free energy can be expanded in terms of the order parameter. The symmetry of the spacetime under the transformation $\varphi \rightleftharpoons -\varphi$ removes the odd powers in that perturbation series. 

\begin{equation}
G(\varphi , T)=G_0 (T) +\frac{1}{2} a(T) \varphi ^2 +\frac{1}{4} b(T) \varphi ^4+...
\end{equation}
where $G_0$ is the Gibbs free energy at $\varphi (T)=0$, which describes the temperature dependence of the high temperature phase near the critical point.

In Landau theory, it is presumed that $b>0$ so that the free energy $G$ has a minimum for finite values of the order parameter $\varphi$. For $a>0$, there is only one minimum at $\varphi =0$, which corresponds to the symmetrical phase (more symmetrical phase). Whereas for  $a<0$ there are two minima with $\varphi\neq 0$ in the unsymmetrical phase (less symmetrical phase). The transition point is governed by the condition of $a=0$. One of the assumptions of the theory is that $a(T)$ has no singularity at the transition point so that it can be expanded in the neighbourhood of the critical point in the integral powers of $(T-T_C)$. To first order,

\begin{equation}
a=a_0(T-T_C) \quad \quad  a_0>0.
\end{equation} 
The coefficient $b(T)$ may also be replaced by $b(T_C)=b$. The expansion of free energy, therefore, becomes,
\begin{equation}
G(\varphi , T)=G_0 (T) +\frac{1}{2} a_0(T-T_C) \varphi ^2 +\frac{1}{4} b \varphi ^4+... \label{Gsecond}
\end{equation}
In the unsymmetrical phase, the dependence of the order parameter $\varphi$ on the temperature near the critical point is determined by the condition that $G$ be minimum as a function of $\varphi$.  In a stable equilibrium state $G(T)$ has a vanishing first derivative and a positive second derivative. 

\begin{eqnarray}
\frac{\partial G}{\partial \varphi}=a_0(T-T_C) \varphi +b \varphi ^3=0 \label{firstd}\\
\frac{\partial ^2 G}{\partial \varphi ^2}=a_0(T-T_C)  +3 b \varphi ^2>0.
\end{eqnarray} 
The solutions of equation \ref{firstd} are,
\begin{equation}
\varphi =0 \quad \text{and} \quad \varphi =\pm \sqrt{-\frac{a_0(T-T_C)}{b}} \label{solution}
\end{equation}

The solution $\varphi =0$ renders a disordered state in the temperature range $T>T_C$ and with $a>0$. The non-zero solution corresponds to the ordered state where the configuration of the phases on the temperature scale depends on the sign of $a_0$. For $a_0>0$ and $a_0<0$, the ordered state corresponds to the temperatures $T<T_C$ and $T>T_C$ respectively. From equation  (\ref{solution}) we have $\varphi \propto (T-T_C)^{1/2}$ near the critical point, which gives the critical exponent $\beta =1/2$.

Substituting these solutions (\ref{solution}) back to the equation \ref{Gsecond} we get the reliance of free energy on the temperature near the phase transition point,

\begin{eqnarray}
G(T,\varphi)=G_0,& & \quad T>T_C\\
G(T,\varphi)=G_0&(T)-\frac{a_0^2}{2b}(T-T_C)^2, &\quad T<T_C.
\end{eqnarray}
These solutions matches at $T=T_C$, i.e., the free energy is continuous at the critical point. At constant pressure the total differential of the Gibbs free energy is 
\begin{equation}
dG=-SdT+\Psi dg +\Pi d\sigma . \label{diffG}
\end{equation}
The expression for entropy is,
\begin{equation}
S=-\left( \frac{\partial G}{\partial T}\right)=\frac{a_0^2}{b} (T-T_C).
\end{equation}
This is the difference of entropy between the ordered and disordered states. If the entropy of the disordered phase is $S_0$, and that of ordered phase is $S_0+\frac{a_0^2}{b} (T-T_C)$. The black hole entropy is also continuous at the phase transition point. The specific heat can be calculated as 

\begin{equation}
C=T\left( \frac{\partial S}{\partial T}\right).
\end{equation}
The specific heat has a jump at the critical point,
\begin{equation}
C(T<T_C)\left| _{T=T_C}-C(T>T_C)\right| _{T=T_C}=\frac{a_0^2}{b}T_C.
\end{equation}
From this it is clear that the heat capacity of the ordered state is greater than that of the disordered state. This expression also indicates that the critical exponent $\alpha$ is zero. From equation \ref{diffG} we have, 
\begin{equation}
-g=\left(\frac{\partial G}{\partial \varphi} \right) _T= a_0(T-T_C) \varphi +b \varphi ^3.
\end{equation}
Which gives,
\begin{equation}
-\left(\frac{\partial \varphi}{\partial g} \right)_T=\frac{1}{a_0(T-T_C)+3b\varphi ^3}. \label{gderivativ}
\end{equation}
Using equation \ref{solution} gives two branches,
\begin{equation}
-\left(\frac{\partial \varphi}{\partial g} \right)_T=\left\lbrace\begin{array}{ll}
\frac{1}{a_0 (T-T_C)} & \text{for} T>T_C\\
\frac{-1}{2a_0 (T-T_C)} & \text{for} T<T_C
\end{array}
\right.
\end{equation}
From this we can infer that the critical exponent $\gamma=1$. At the phase transition point $a=0$, therefore, from equation \ref{gderivativ} we can obtain the relation,
\begin{equation}
g\propto \varphi ^3.
\end{equation}
Which simply tells that the critical exponent $\delta =3$.


\section{Ruppeiner Geometry and Microstructure}
\label{sectwo}
\begin{figure}[H]
\centering
\includegraphics[scale=1]{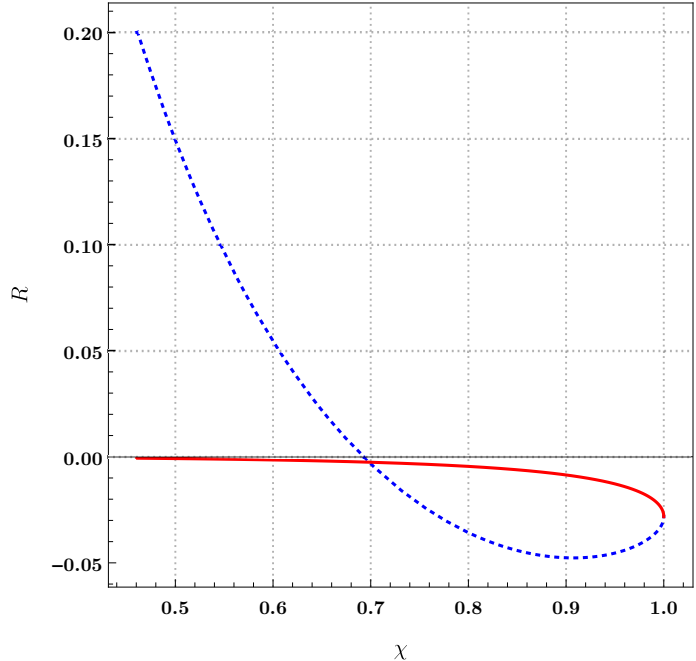}
\caption{The Ruppeiner scalar curvature vs reduced temperature $(R - \chi)$ plot for regular Hayward AdS black hole.}\label{geometrycurve}
\end{figure}

Landau theory, being universal, does not give the phase structure of the system. However, the symmetry or order of degree of the system arises from the underlying microstructure. The parameters $a$ and $b$ in the theory are related to the system characteristics but they do not appear in the identification of the critical exponents. This discrepancy persists even for an ordinary thermodynamic system. This is because, in the continuous phase transition theory the fluctuation of the order parameter is not considered near the transition point. This can be addressed by using a fluctuation theoretical tool namely the Ruppeiner geometry. The black hole microstructure can be analyzed by studying the nature of Ruppeiner curvature scalar $R$. The thermodynamic invariant curvature scalar $R$ is calculated by using the definition in the Weinhold energy form \citep{weinhold1975metric},
\begin{equation}
g_{\alpha \beta}=\frac{1}{T}\frac{\partial ^2M }{\partial  X^\alpha \partial  X^\beta}
\end{equation}
in the $(S,P)$ parametric space as
\begin{equation}
R=\frac{5 \pi ^{3/2} g^3 \sqrt{S}-S^2}{\left(\pi ^{3/2} g^3+S^{3/2}\right) \left(S^{3/2} (8 P S+1)-2 \pi ^{3/2} g^3\right)}.
\end{equation}
Using the earlier results (equations \ref{r2eqn}, \ref{peqn} and \ref{teqn}) we obtain the curvature scalar for the two phases of the black hole. The obtained $R$ is plotted against $\chi$ to study the nature of interaction between the black hole molecules (fig. \ref{geometrycurve}). In the magnetically charged black hole, there are two distinct types of molecules; uncharged and charged, which contributes to the microscopic degrees of freedom of the entropy. The black hole phase transition can be seen as the manifestation of the phase structure of a two-fluid system with magnetically charged and uncharged molecules.

The ordered and disordered phase of the black hole can be attributed to the relative degree of freedom of the magnetically charged molecules. If the d.o.f. of charged molecules is $N_g$ and the total d.o.f the black hole is $N$, the black hole can have three different situations depending on $N_g/N$ ratio. $N_g/N=n_0$ the moderately ordered, $N_g/N>n_0$ highly ordered and $N_g/N<n_0$ less ordered phases, due to the action of the magnetic potential $\Psi$. 

At a given temperature the magnetic potentials corresponding to the $R_1$ and $R_2$ branches are different. Since $\Psi _2 <\Psi _1$, $R_2$ represents the symmetric phase of the black hole where the molecules are in a disordered state. In other words $R_2$ stands for the low potential phase of the black hole. Therefore $R_2$ branch always represents the situation $N_g/N<n_0$. Whereas the phases corresponding to $R_1$ have three different cases. $R_1>0$ phase $(N_g/N<n_0)$ has less symmetry and higher order, $R_1=0$ phase $(N_g/N=n_0)$ has both moderate symmetry and order, $R_1<0$ phase $(N_g/N>n_0)$ has more symmetry and lower order.

\section{Discussions}
\label{secthree}
From several recent developments in the field of black hole thermodynamics, it is a well-established notion that black holes have microstructure like ordinary thermodynamic systems. And the degrees of freedom of these constituent black hole molecules are what counts for the black hole entropy. The phase structure of these molecules entitles the thermodynamic and phase transition properties of the system. In this work, we have shown that the phase transition of a magnetically charged black hole is determined by the magnetic potential. However this result is similar to that of RN-AdS black hole where the key role is played by the electric potential \citep{Guo2019}. The symmetry or the order degree of the regular black hole is governed by the magnetic potential. The changing symmetry results in different phases of the system, which is investigated using the Landau theory of continuous phase transition. 

The statistical interpretation of the Ruppeiner scalar reveals the nature of the interaction of the black hole molecules. Under the limiting case $g\rightarrow0$ we have $R<0$, which corresponds to uncharged molecules. With the presence of charged molecules, $R$ tends to become positive. In the context of quantum gases $R<0$ and $R>0$ results are obtained for fermion gas and boson gas, respectively \citep{Mirza2008, PhysRevE.88.032123}. For the regular Hayward black hole, magnetically charged molecules and uncharged molecules have different microstructures similar to fermion gas and boson gas. 

However, in the case of anyon gas, the sign of $R$ tells the average interaction between the constituents. Positive $R$ stands for repulsive and negative $R$ corresponds to attractive interaction. Vanishing $R$ implies zero interaction. In this view, we can think the black hole molecules have repulsive, attractive and zero interactions as per the curvature scalar behaviour. Our research gives information about the microstructure of magnetically charged molecules which is similar to the electrically charged molecules in RN-AdS black hole. This phenomenological description will help us to understand the exact microstructure of the black hole in the future research. The similar investigation can be done for other non singular black holes.


\acknowledgments
Author N.K.A. would like to thank U.G.C. Govt. of India for financial assistance under UGC-NET-SRF scheme.


  \bibliography{BibTex}

\end{document}